\documentclass{osa-article}
\usepackage{mymacros}
\usepackage{physics}

\journal{osajournal}

\articletype{Research Article}

\begin{document}

\title{Quantum detector tomography of superconducting nanostrip photon-number-resolving detector}

\author{Mamoru Endo,\authormark{1,4} Tatsuki Sonoyama,\authormark{1} Mikihisa Matsuyama,\authormark{1} Fumiya Okamoto,\authormark{1} Shigehito Miki,\authormark{2,3}  Masahiro Yabuno,\authormark{2} Fumihiro China,\authormark{2} Hirotaka Terai,\authormark{2} and Akira Furusawa\authormark{1,5}}

\address{\authormark{1}Department of Applied Physics, School of Engineering, The University of Tokyo, 7-3-1 Hongo, Bunkyo, Tokyo 113-8656, Japan\\
\authormark{2}Advanced ICT Research Institute, National Institute of Information and Communications Technology, 588-2 Iwaoka, Nishi-ku, Kobe, Hyogo 651-2492, Japan\\
\authormark{3}Graduate School of Engineering, Kobe University, 1-1 Rokkodai-cho, Nada-ku, Kobe-city, Hyogo 657-0013, Japan\\
\authormark{4}endo@ap.t.u-tokyo.ac.jp\\
\authormark{5}akiraf@ap.t.u-tokyo.ac.jp}



\begin{abstract}
Superconducting nanostrip photon detectors have been used as single photon detectors, which can discriminate only photons' presence or absence. It has recently been found that they can discriminate the number of photons by analyzing the output signal waveform, and they are expected to be used in various fields, especially in optical quantum information processing. Here, we improve the photon-number-resolving performance for light with a high-average photon number by pattern matching of the output signal waveform. Furthermore, we estimate the positive-operator-valued measure of the detector by a quantum detector tomography. The result shows that the device has photon-number-resolving performance up to five photons without any multiplexing or arraying, indicating that it is useful as a photon-number-resolving detector.
\end{abstract}

\section{Introduction}
Photon detectors, which can count photons, are widely used in many fields, including quantum optics \cite{Hadfield2009}, metrology \cite{Shangguan2017}, or imaging \cite{Morimoto2020}, where ultra-weak light detection is required. Typically, photomultiplier tubes (PMTs) \cite{Yamazaki1985}, avalanche photodiodes (APDs) \cite{Cova1981}, superconducting nanostrip photon detectors (SNSPDs) \cite{Goltsman2001}, and superconducting transition edge sensors (TESs) \cite{Cabrera1998} are in use. Most of these detectors, except TESs or other special detectors \cite{Kardyna2008} , are single-photon detectors that only determine the presence or absence of photons. In contrast, photon-number-resolving detectors (PNRDs) such as TESs, which can discriminate even the number of photons, are desired as ideal detectors in myriad fields \cite{Gerrits2010,Namekata2010,Giustina2015, Niwa2017}. 

Very recently, the quantum computational advantage with photons has been demonstrated by Gaussian boson sampling, where 100 single-photon detectors based on SNSPDs were used \cite{Zhong2020}. However, since realistically available detectors are limited to single-photon detectors, one can perform only approximate calculations. Replacing the single-photon detectors with PNRDs would essentially advance such research. Besides, in continuous-variable quantum information processing \cite{Lloyd1999,Andersen2015,Takeda2019}, it is essential to generate a special quantum state called non-Gaussian quantum states of light toward fault-tolerant universal quantum computer, where a PNRD is a key component \cite{Gottesman2001,Alexander2018}. Although non-Gaussian quantum states have been actively studied using various types of photon detectors \cite{Namekata2010,Gerrits2010,Yukawa2013,Asavanant2017, Ra2020a}, PNRDs that can be easily incorporated into real experimental systems are highly demanded.
 
There are two main ways to realize a PNRD. One is to multiplex single-photon detectors spatially or temporally. Geiger mode APDs \cite{Jiang2007, Morimoto2020} or SNSPDs \cite{Divochiy2008, Natarajan2013, Gaggero2018, Yabuno2020} are often used. These methods achieve compact photon-number-resolving performance by laying out the sensor in a small area, which are quite versatile for many applications. The other way is to use a single element sensor with photon-number-resolving performance, such as a TES, which can discriminate the photon number with almost unity detection efficiency \cite{Fukuda2011b, Lita2008}.

However, it is known that to realize an ideal PNRD by the method of multiplexed single-photon detectors is challenging task from theoretical considerations, especially in optical-quantum-information-processing applications, where one must need to know the exact photon number \cite{Sperling2012,Jonsson2019,Bartlett2002,Provaznik2020}. This can be explained based on a positive-operator-valued-measure (POVM) formalism, which mathematically characterize a non-projective quantum detectors such as PNRDs (the details can be found in a later section)\cite{Provaznik2020}. If we know the POVM elements $\{\pi_n\}$ of a detector, we can calculate the output statistics of the detector with respect to the input as $p_{n,\rho}=\text{tr}[\rho\pi_n]$, where $p_{n,\rho}$ is the output statistics of obtaining detection outcome $n$ when a quantum state $\rho$ is input. Also, by measuring the POVM of a detector, we can determine how close the detector is to the perfect PNRD. For example, let us consider a target detector to measure the photon number state (also known as Fock state) $\ket{n}$. In the case of a perfect PNRD, when the detector output is $n$, the input state can be determined to be $\ket{n}$ with 100\% probability, that is, $\pi_n=\dyad{n}$. Surprisingly, the probability that the input state is $\ket{n}$ is the highest when the detector output is $n$, even if the detection efficiency of the PNRD is somewhat low \cite{Provaznik2020}. In contrast, in the case of  multiplexed single-photon detectors to mimic an PNRD, an unrealistic number of single-photon detectors with almost 100\% detection efficiency must be multiplexed to achieve the same result of the single-element PNRD mentioned above. In practical, the PNRD required for optical-quantum-information processing (i.e., a PNRD which can assert that the input state is $\ket{n}$ when the output is $n$) is only realized by a single-element PNRD. Furthermore, time-multiplexed configuration may limit detection speed \cite{Kruse2017}, and spatial array increases the dark count and the system complexity.  Although, in recent years, the development of SNSPDs has enabled us to achieve almost 100\% detection efficiency with negligible dark count \cite{Reddy2020},  large-scale multiplexing involves many technical challenges. 

Due to this background, especially in optical quantum information processing, TESs have been almost exclusively regarded as an ideal PNRD. However, the biggest challenge of TESs is its time performance. While most single-photon detectors have a temporal resolution of less than 100 ps \cite{Goltsman2001,Korzh2020}, TESs are approximately 100 times worse \cite{Lamas-Linares2013}. Also, TESs require a cryogenic environment of about 100 mK, which requires a large refrigeration system such as a dilution refrigerator or adiabatic demagnetization refrigerator (ADR). In addition, a superconducting quantum interferometer is typically required to read out the signal. For these reasons only a few research groups can install TESs.

Recently, it has been reported that an SNSPD can detect up to four photons based on the slew rate \cite{Cahall2017} or the height of the signals waveform \cite{Zhu2020}. The commonly used readout circuit for an SNSPD can not show the difference in signals due to the photon number, but by using a low-noise amplifier or adding a special structure to the SNSPD, it becomes possible to distinguish the small difference in signal due to the number of photons. But the performance as the quantum detector of such a photon-number-resolving SNSPD has not been clarified. Here, we focus on the POVM as a quantity that fully characterizes the SNSPD-based PNRD. We used a technique called detector tomography to estimate the POVM \cite{Lundeen2009}. The results show that the sensor can determine the photon number with high accuracy for as few as five photons, consistent with the theoretically expected value. Furthermore, since our method uses the waveform information to discriminate the number of photons, the identification performance is improved. It will be easy to combine this method with a high-speed digitizer and a field-programmable gate array (FPGA), and high-speed, real-time photon-number resolving will be possible.

\section{Superconducting nanostrip photon-number-resolving detector}
\begin{figure}[h!]
\centering\includegraphics[width=13cm]{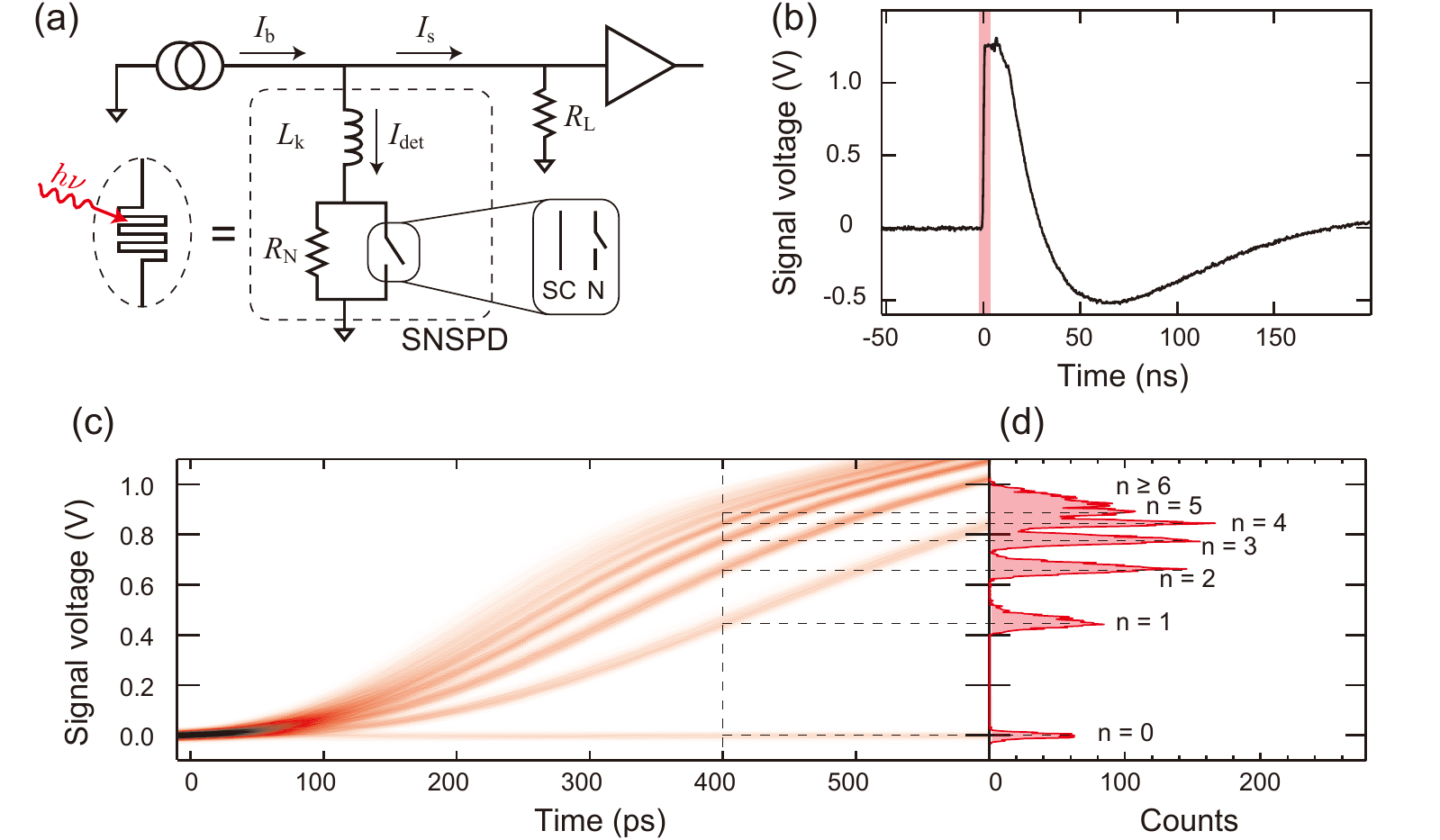}
\caption{(a) Electrical model of an SNSPD and a simplified readout circuit. (b) An example of oscilloscope trace of single-shot photon-detection event with a readout circuit shown in Fig.~\ref{fig:setup}.  (c) Overlay of 10,000 traces around the red region in (b), when $|\alpha|^2=5.7$. The SNSPD discriminates input photon number up to five. (d) Waveform histogram at the time of 400 ps (vertical dashed line in (c)). Each peak represents the number of photons ($n=0,\dots,5$ and $n\ge6$).} 
\label{fig:model_waveforms}
\end{figure}
By cooling an SNSPD to well below the critical temperature and applying a bias current slightly lower than the critical current, the SNSPD was shown to be sensitive to single photons \cite{Goltsman2001,Natarajan2012a}. When a photon with an energy ($h\nu$) greater than the superconducting gap is incident on it, a normal-conductive spot called a hotspot is created. The creation process of the hotspot is a physically interesting subject \cite{Renema2014a, Bulaevskii2012}, but since it is sufficiently faster than the time scale considered in this research, we assume that the hotspot is created instantaneously and focus on the phenomena after it is created. The superconducting current flows in such a way as to avoid the hotspot, but since the nanostrip is narrow (typically narrower than 100 nm), the current density around the hotspot increases and exceeds the critical current. This continues until the hotspot spreads over the entire width of the nanostrip, creating a resistive barrier. The temperature of the nanostrip then drops because the current is no longer flowing, and it returns to its initial state. When multiple photons are injected within a relaxation time of the hotspot (typically the order of 100~ps \cite{Heeres2012,Marsili2016}), hotspots are created simultaneously along the nanostrip. Therefore, the resistance of the nanostrip depends on the number of photons incident on the nanostrip. In other words, a single-element SNSPD acts as a highly multiplexed single-photon detector.

\begin{figure}[h!]
\centering\includegraphics[width=13cm]{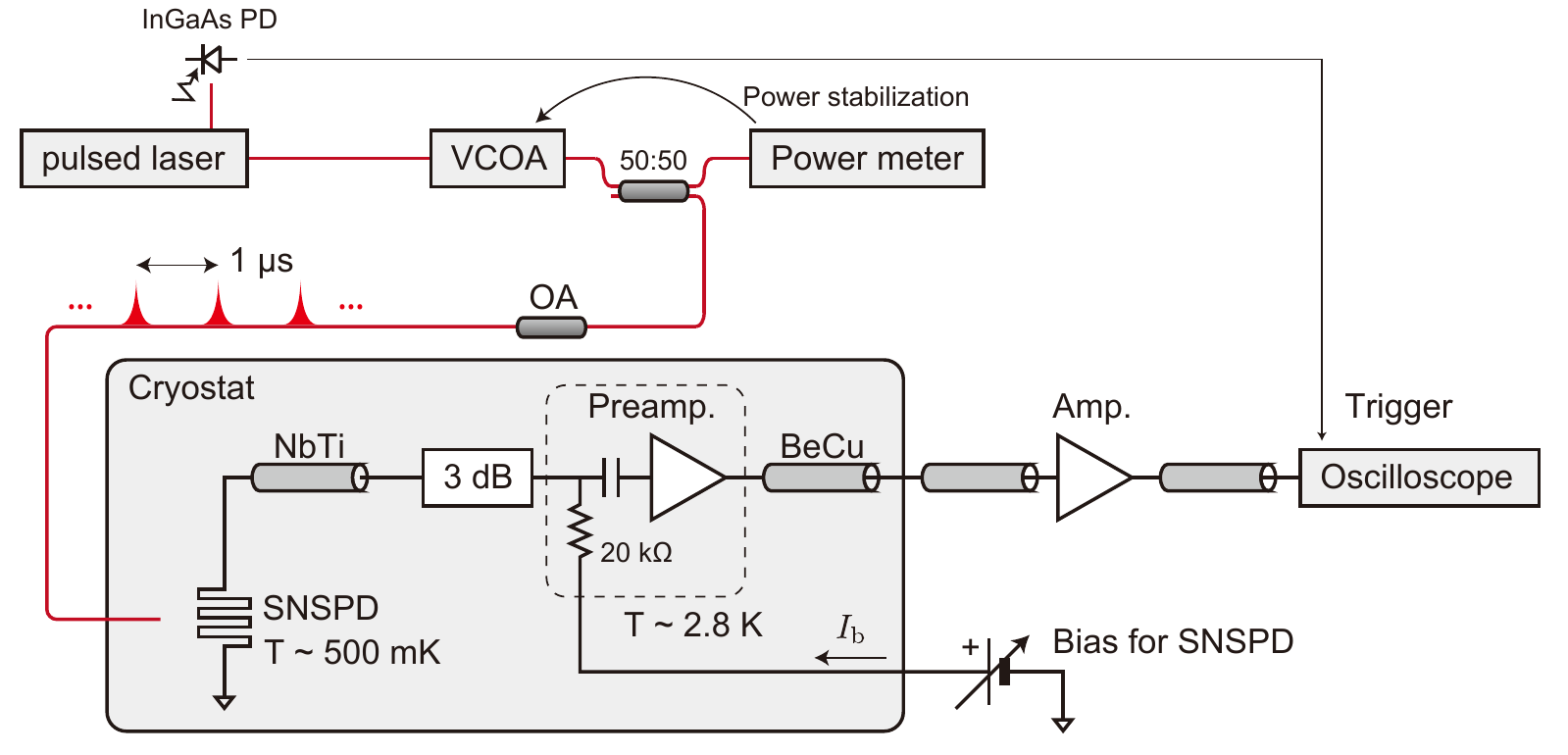}
\caption{Experimental apparatus. A power-calibrated, pulsed coherent light is coupled to the SNSPD in the cryostat. The SNSPD is cooled down to around 500 mK. InGaAs PD: InGaAs photodiode for triggering the oscilloscope, VCOA: voltage-controlled optical attenuator, OA: calibrated optical attenuator, NbTi: Niobium-titanium coaxial cable, 3 dB: 3-dB attenuator, Preamp.: cryogenic amplifier and a bias tee with a series resistance of 20 k$\Omega$, BeCu: Beryllium copper coaxial cable, Amp.: booster amplifier.} 
\label{fig:setup}
\end{figure}

The electrical model consists of a kinetic inductance ($L_\text{k}$), a time-dependent resistance ($R_\text{N}$), and a switch \cite{Natarajan2012a, Nicolich2019}, as shown in Fig.~\ref{fig:model_waveforms}(a). The bias current ($I_\text{b}$) is supplied by a current supply (typically a voltage supply with a series resistor) and splits between the current passing through the SNSPD ($I_\text{det}$) and the signal current ($I_\text{s}$) passing through the load resistor of readout circuit ($R_\text{L}$, usually 50~$\Omega$). Initially, the switch is closed (SC: superconducting state), and most of the bias current flows through the switch. When a photon is injected, a hotspot is created, the switch opens (N: normal state), and the current flows through the resistor, resulting in a voltage signal. Figure~\ref{fig:model_waveforms}(b) shows an example of signal waveform when the SNSPD is irradiated with a picosecond pulsed laser by the readout circuit as shown in Fig.~\ref{fig:setup}. The process from the creation of the hotspot to its disappearance corresponds to the rising part of the signal waveform (red occupied). At this time, the resistance value changes in accordance with the number of incident photons, and thus the signal wave height and rise time show dependence on the number of photons. Note that the undershooting of the waveform at the falling edge may be due to impedance mismatch between circuits at cryogenic temperatures, but this is not a problem in this study.

Figure~\ref{fig:model_waveforms}(c) shows that the overlay of 10,000 waveforms is shown when a picosecond optical pulse with $|\alpha|^2=5.7$ is injected. As shown in the figure, the rising part of the waveform differs depending on the number of photons. The histogram at 400~ps is shown in Fig.~\ref{fig:model_waveforms}(d). It can be seen that there is a peak corresponding to each photon number, indicating that up to five photons can be resolved. Note that the signal-to-noise ratio of a typical SNSPD readout circuit is not sufficient, and it is not possible to see the difference in the rising edge. The previous studies suggest using an ultra-low noise amplifier at cryogenic temperatures for photon-number resolving \cite{Cahall2017}. Also, the selection of coaxial cables, amplifiers used in the later stages, and an oscilloscope is quite important.

\section{Experimental details}
\subsection{Experimental apparatus}
The experimental apparatus is shown in Fig.~\ref{fig:setup}. The light source is a frequency-doubled mode-locked Er:fiber laser (Menlo Systems, M-Comb). The center wavelength is 772.66 nm with a pulse duration of approximately 1 ps , which is much shorter than the relaxation time of the hotspot \cite{Marsili2016}, and a repetition rate of 1 MHz, after an optical bandpass filter, Er-doped fiber amplifiers, a pulse picker based on an acousto-optic modulator, and a second-harmonic module (not shown in the figure). A small portion of light at the fundamental wavelength is detected by a fast InGaAs photodiode for a trigger of an oscilloscope. The residual light is attenuated by a voltage-controlled optical attenuator (VCOA) and then split by a 50:50 fiber beam splitter to control and monitor the optical power. For power stabilization, a home-made feedback circuit is used, and the optical power is stabilized within 1\%. The other output of the fiber beam splitter is attenuated again by a calibrated optical attenuator (OA) to reach the single photon level. 
The detector is a fiber-coupled SNSPD with 100-nm-wide niobium-titanium-nitride (NbTiN) nanostrip spaced on a 180-nm pitch, meandering over a $15\times15\ \mu\text{m}^2$ active area. The device has a dielectric multilayer cavity to enhance its detection efficiency \cite{Yamashita2016}. Note that this SNSPD is designed for the wavelength of 860 nm, not for 772.66 nm, so the detection efficiency is not optimized for this experiment.
The details of this device can be found in the references \cite{Miki2013,Miki2010}. A standard single-mode fiber (Nufern, 780HP) is used as a delivering fiber. The SNSPD is operated at a Gadolinium-Gallium-Garnet stage in an ADR (ENTROPY, ADR-M) in the temperature range of 500~mK. The cryogenic amplifier (Cosmic Microwave Technologies, CITLF3-20K) is used as a preamplifier with the operating temperature of 2.8~K. This amplifier has a bias tee with a series resistance of 20 k$\Omega$ for biasing the SNSPD. The SNSPD and the preamplifier is connected with a superconducting coaxial cable (COAX Co., Ltd., SC-086/50-NbTi-NbTi). Note that before the preamplifier, a 3-dB attenuator (Mini-circuits, BW-S3-2W263+) is installed, which mitigates current back-action at high count rates \cite{Kerman2013,Cahall2018}. A Beryllium copper coaxial cable (COAX Co., Ltd., SC-086/50-B-B) is used to connect the preamplifier and a feedthrough of the ADR. Outside the ADR, an additional low-noise amplifier (PASTERNAK, PE15A1007) boosts the output signal. The signal is monitored and recorded by a high-speed oscilloscope (Tektronix, MSO6B, 10-GHz analog bandwidth, 50~Gigasamples/s). The bias current of the SNSPD is set to be 33.5~$\mu$A during the experiment. At that current, a dark count is less than one count per second and the measured detection efficiency is 54.7\%. The bias current supply for the SNSPD and the power supply for the preamplifier are home-made low-noise battery power supplies to avoid noises due to ground loops. Note that typically the SNSPD’s detection efficiency is dependent on the input polarization \cite{Natarajan2013}, but in our case, the dependence was not very critical, so we do not optimize the input polarization. 

\begin{figure}[h!]
\centering\includegraphics[width=13cm]{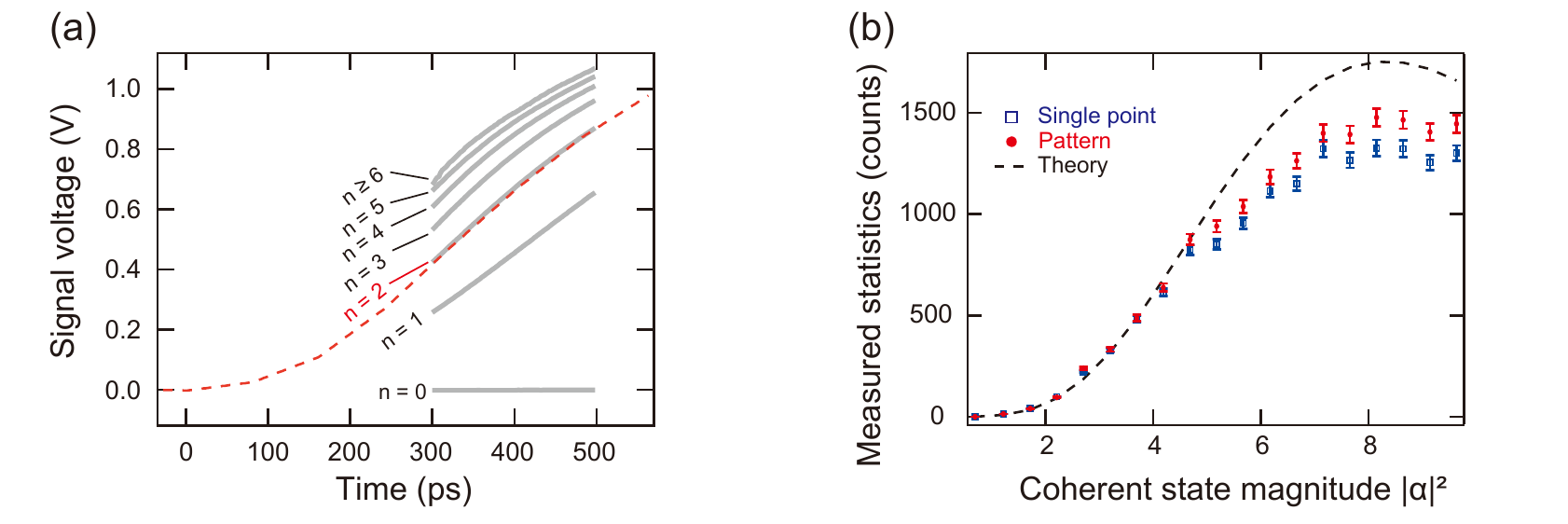}
\caption{(a) Photon number discrimination by pattern matching. Gray lines show reference traces, which correspond to the photon number of $n = 0,\dots,5$ and $n\ge6$.  The dashed red line shows an example waveform with $n = 2$. (b) Comparison of measured statistics between the proposed method (red circle, pattern), the single-point method (blue square, single point), and the theoretical values (black dashed line, theory), when $n=5$. } 
\label{fig:pattern}
\end{figure}
\subsection{Photon number discrimination by waveform pattern matching}
In the first attempt on photon-number resolving with an SNSPD \cite{Cahall2017}, the electrical signal was passed through an electrical differentiation circuit, and the photon number was discriminated from the signal height value. In this paper, we prepared a reference waveform corresponding to each photon number in advance and determined the photon number by calculating which reference waveform is the closest to an acquired waveform. The advantage of this method is that the photon number is determined based on the waveform information, not just the information at a certain point, so it is highly resistant to noise. This is especially effective for waveforms for a large number of photons because the overlap between waveforms is large and difficult to distinguish. 

To obtain reference waveforms, we injected probe light into the SNSPD with the coherent state magnitude of $|\alpha|^2 = 5.7$ and acquired 10,000 waveforms. The peak of the histogram at each time (also see Fig.~\ref{fig:model_waveforms}(d)) was calculated and connected to determine the waveform corresponding to each photon number. Here, we classify the photons into 0, 1, 2, 3, 4, 5 photon(s), and 6 or more photons (gray lines in Fig.~\ref{fig:pattern}(a)). The light to be measured is injected, and the waveform is acquired as shown in a dashed red line in Fig.~\ref{fig:pattern}(a). The sum of the squared difference between the target waveform and the reference waveform for each photon number is calculated, and the one reference with the smallest value is selected. In this figure, the reference for $n=2$ is the closest for example. The range of the waveform used for pattern matching is between 300 ps and 500 ps. Using a wider range of waveform information did not improve the discrimination performance. When $n$ is small, there is no difference from the result with photon-number discrimination using information from only single point (here we use the one signal value at 400 ps as shown in Fig.~\ref{fig:model_waveforms}(d)). It is  because the noise of the instruments is sufficiently small in our experimental apparatus. However, when the number of photons becomes large and the input power becomes high, such as $n=5$ and $|\alpha|^2> 4$, the proposed method is closer to the theoretical line as shown in Fig.~\ref{fig:pattern}(b). The power meter we used is traceable to Physikalisch-Technische Bundesanstalt (PTB), but has a measurement uncertainty of 3\%, which is the dominant error in our experiment. The error bars in Fig.~\ref{fig:pattern}(b) correspond to this uncertainty.

In this paper, the pattern matching was performed on a computer after all waveforms were acquired by the oscilloscope. This method does not require complex calculations and can be performed in real time and with low latency if a sufficiently fast digitizer and an FPGA are available. This means that not only real-time measurement but also operations based on photon numbers can be realized, which is a key technique for implementing optical quantum information processing.

\begin{figure}[h!]
\centering\includegraphics[width=13cm]{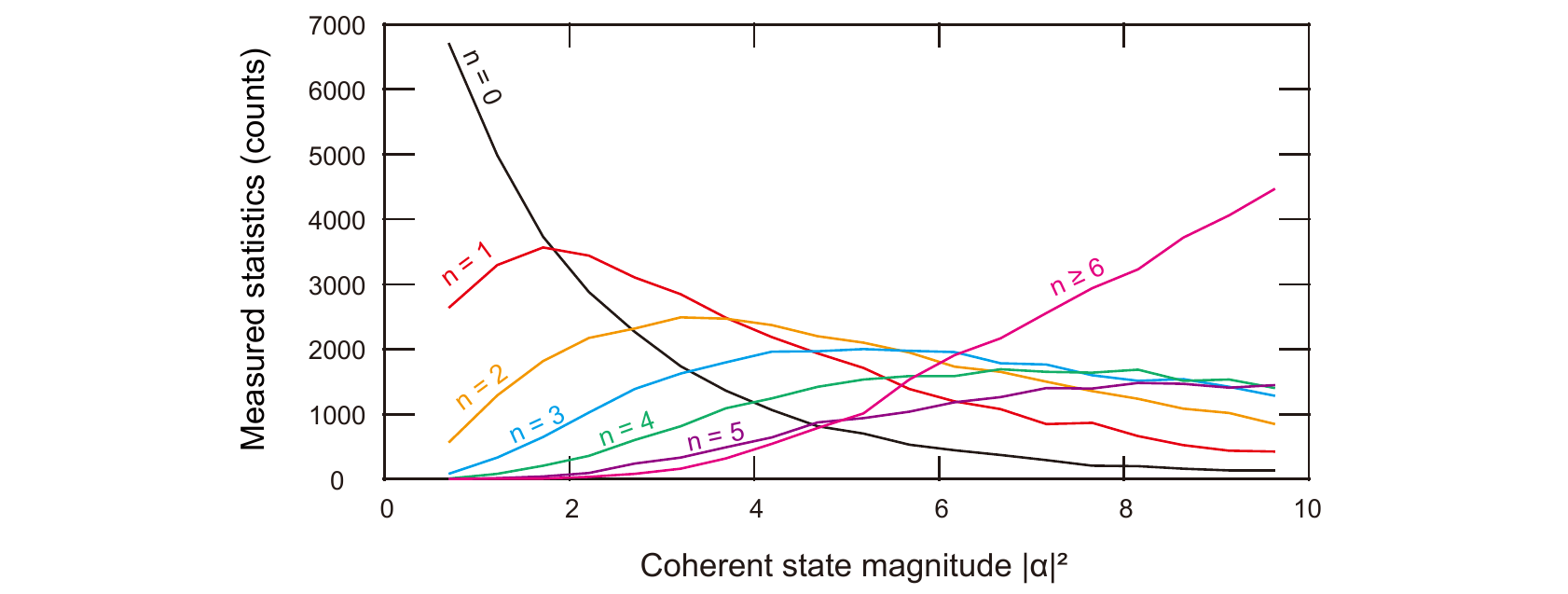}
\caption{(c) Measured statistics for different $n$.  } 
\label{fig:statistics}
\end{figure}
\section{Results}
\subsection{Detector tomography}
Non-projective detectors such as PNRDs can be described mathematically using the POVM formalism \cite{DAriano2004,Luis1999}. If the POVM elements $\{\pi_n\}$ of a detector are known, the detection probability $p_{n, \rho}$ of obtaining detection outcome $n$ when a state $\rho$ is input to the detector can be obtained as 
\begin{align}
p_{n,\rho}=\text{tr}[\rho\pi_n],
\end{align}
where $\pi_n \ge 0$ and $\sum_n\pi_n=I$ for a physical detector. Conversely, if the response of a detector to a set of a sufficient number of probe states is known, the POVM element can be estimated by inverse calculation. This method is called as detector tomography \cite{Lundeen2009,Natarajan2013,Yokoyama2019}. Coherent states $\ket{\alpha}$, which are overcomplete in Hilbert space for an optical mode, are suitable as probe states. For detectors that are insensitive to the phase of the state, such as a PNRD, the number of probe states required for tomography can be greatly reduced because the POVM depends only on the amplitude of the coherent state $|\alpha|^2$. In other words, we only need to prepare coherent states with different amplitudes and examine the output of the PNRD to be measured. The phase-insensitive detector can be represented by the diagonal component of the POVM when expressed in the photon number basis as 
\begin{align}
\pi_n=\sum_{k=0}^\infty \theta_k^{(n)}\dyad{k}.
\end{align}
Note that, for an ideal PNRD with a unity detection efficiency, $\pi_n$ equals to $\dyad{n}$, that is $\theta _k^{(n)} = \delta_{n,k}$. For an $N$-outcome detector, the matrix representation can be used when truncating to $M$ number states as $P_{D\times N}=F_{D\times M}\Pi_{M\times N}$, where $P_{D\times N}$ is measurement statistics for $D$ probe states, $F_{D\times N}$ contains the $D$ probe states $\{\alpha_1,\dots\alpha_D\}$, and $\Pi_{M\times N}$ is the POVM which is wanted to know (subscripts show matrix dimensions).   For coherent state probes, $F_{i,k}=\left[|\alpha_i|^{2k}\exp(-|\alpha_i|^2)\right]/k!$. As a result, the POVM element can be estimated by solving the following optimization problem, 
\begin{align}
\text{min}\{||P-F\Pi||_2+g(\Pi)\},\ \ \text{subject to}\ \ \pi_n\ge 0, \sum_{n=0}^{N-1}\pi_n=I,
\end{align}
 where $||A||_2$ is defined as $\sqrt{\sum_{i,j}|A_{i,j}|^2}$, and $g(\Pi)=\gamma\sum_{k,n}\left[\theta_k^{(n)}-\theta_{k+1}^{(n)}\right]^2$ is a function to calculate the sum of squares of the differences of adjacent POVM elements. In this paper, we use $\gamma = 0.01$. The details can be found in the references \cite{Lundeen2009}.

\begin{figure}[h!]
\centering\includegraphics[width=13cm]{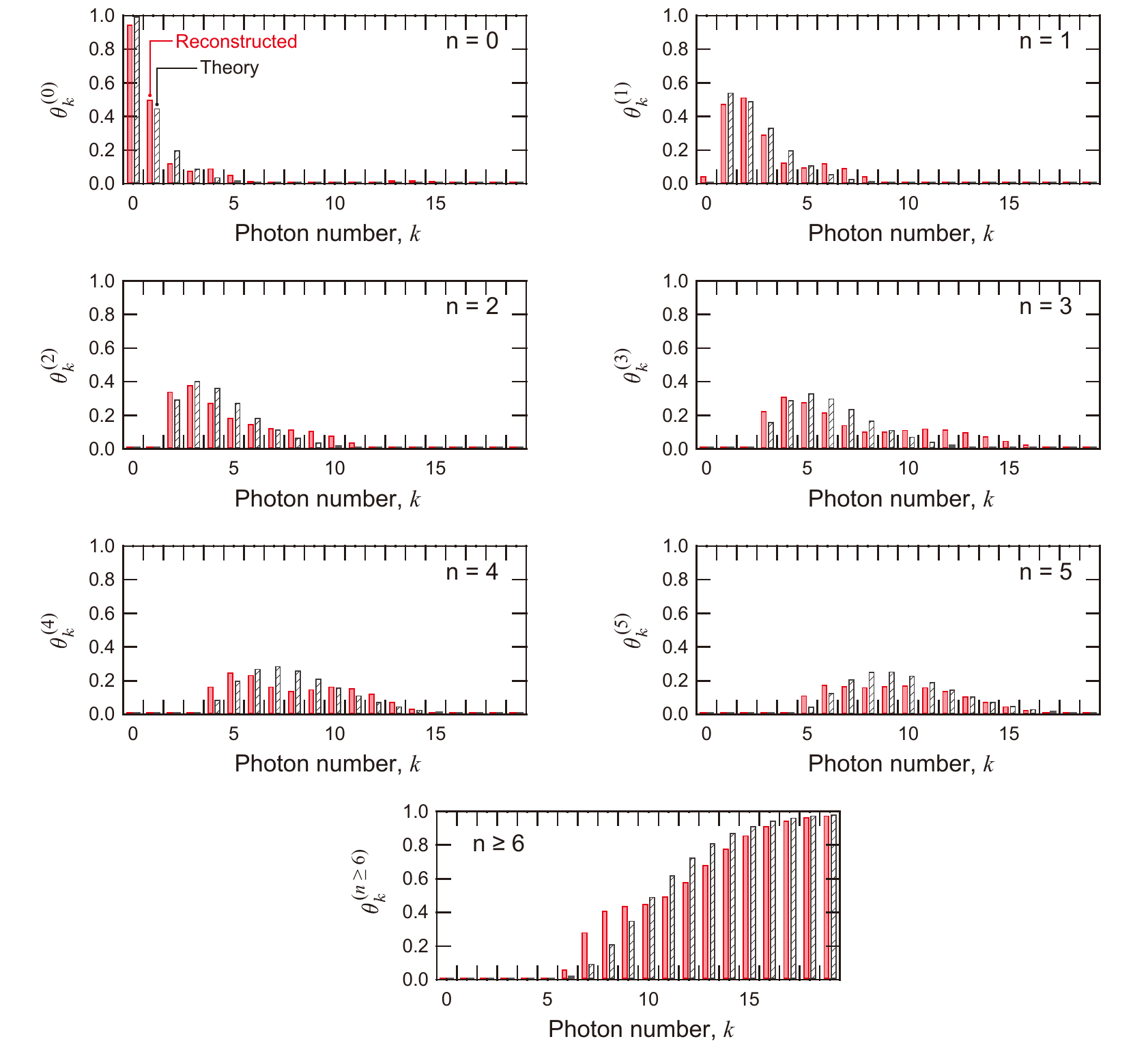}
\caption{The diagonals of the reconstructed  POVM elements $\theta_k^{(n)}$ for the photon-number-resolving SNSPD (red occupied). The theoretical POVM elements for a PNRD with the detection efficiency of 54.7\% are also shown (black diagonal line). } 
\label{fig:povm}
\end{figure}
\subsection{POVM estimation}
The measured statistics for the number of photons on the SNSPD is plotted in Fig.~\ref{fig:statistics}, where the number of coherent probe states was $D=19$, and 10,000 waveforms were acquired for each probe. The reconstructed POVM elements $\theta_k^{(n)}$ for $n=0,\dots,5$ and $n\ge 6$, that is $N=7$, based on the measured statistics are shown in Fig.~\ref{fig:povm}. Note that we choose $M=70$ for the POVM estimation. Again, we assume the off-diagonal POVM elements are zero. Theoretical values for a PNRD with a 54.7\% detection efficiency are also plotted in the figure \cite{Achilles2004}.  Because the dark counts of the SNSPD used in this study are negligible, the POVM element for $k< n$ are close to zero. The discrepancy between the estimated and theoretical values is due to the aforementioned errors in the power meter and the incompleteness of the SNSPD. This may be due to the fact that we set $\gamma = 0.01$ when we reconstructed the POVM, but this value may not be appropriate. Also, a more accurate POVM can be estimated if a large number of probes can be used to perform the detector tomography.

\section{Conclusion}
By discriminating the photon number from the signal waveform of the SNSPD, we measured the POVM using the detector tomography technique and clarified its performance as a quantum detector. The resolving performance is increased for higher photon numbers by applying the simple pattern matching method. Compared to TESs, the maximum measurable photon number itself is inferior, but the SNSPD operates about two orders of magnitude faster. In this experiment, the overall detection efficiency is limited to 54.7\%, because the SNSPD used here is not designed for the probe wavelength of 772.66~nm. Also the losses of fiber splicing and connectors are present. Recently, an SNSPD with almost unity detection efficiency has been developed\cite{Reddy2020}, and by using such devices and reducing other losses, a versatile PNRD can be realized. Since it is possible to add photon-number-resolving performance to an SNSPD, simply by adding a cryogenic amplifier, it can be easily expanded when an SNSPD is already installed. In particular, in the field of optical quantum information processing, single-photon detectors, which discriminate only the presence or absence of photons, are fundamentally different from PNRDs, and this should allow us to dramatically improve experiments that could only be performed approximately so far \cite{Zhong2020}.

\section*{Fundings}
Japan Science and Technology Agency Moonshot R\&D (Grant No. JPMJMS2064), Japan Society for the Promotion of Science KAKENHI (Grant No. 18H05207 and 20K15187).
\section*{Acknowledgments}
The authors acknowledges supports from UTokyo Foundation, and donations from Nichia Corporation of Japan.  M.E. acknowledges supports from Research Foundation for Opto-Science and Technology. T.S. and M.M. acknowledge supports from The Forefront Physics and Mathematics Program to Drive Transformation (FoPM). F.O. acknowledges supports from Advanced Leading Graduate Course for Photon Science (ALPS). The authors would like to thank Dr. Warit Asavanant for fruitful discussions. The authors would like to thank Mr. Takahiro Mitani for careful proofreading of the manuscript. The authors would like to thank Tektronix, Inc. for loaning the high-speed oscilloscope (MSO6B).

\section*{Disclosures}
The authors declare no conflicts of interest.


\bibliography{snpnrd_povm}

\begin{thebibliography}{10}
\newcommand{\enquote}[1]{``#1''}

\bibitem{Hadfield2009}
R.~H. Hadfield, \enquote{{Single-photon detectors for optical quantum
  information applications},} {\protect\JournalTitle{Nature Photonics}}
  \textbf{3}, 696--705 (2009).

\bibitem{Shangguan2017}
M.~Shangguan, H.~Xia, C.~Wang, J.~Qiu, S.~Lin, X.~Dou, Q.~Zhang, and J.-W. Pan,
  \enquote{{Dual-frequency Doppler lidar for wind detection with a
  superconducting nanowire single-photon detector},}
  {\protect\JournalTitle{Optics Letters}} \textbf{42}, 3541 (2017).

\bibitem{Morimoto2020}
K.~Morimoto, A.~Ardelean, M.-L. Wu, A.~C. Ulku, I.~M. Antolovic, C.~Bruschini,
  and E.~Charbon, \enquote{{Megapixel time-gated SPAD image sensor for 2D and
  3D imaging applications},} {\protect\JournalTitle{Optica}} \textbf{7},
  346--354 (2020).

\bibitem{Yamazaki1985}
I.~Yamazaki, N.~Tamai, H.~Kume, H.~Tsuchiya, and K.~Oba,
  \enquote{{Microchannel-plate photomultiplier applicability to the
  time-correlated photon-counting method},} {\protect\JournalTitle{Review of
  Scientific Instruments}} \textbf{56}, 1187--1194 (1985).

\bibitem{Cova1981}
S.~Cova, A.~Longoni, and A.~Andreoni, \enquote{{Towards picosecond resolution
  with single‐photon avalanche diodes},} {\protect\JournalTitle{Review of
  Scientific Instruments}} \textbf{52}, 408--412 (1981).

\bibitem{Goltsman2001}
G.~N. Gol'tsman, O.~Okunev, G.~Chulkova, A.~Lipatov, A.~Semenov, K.~Smirnov,
  B.~Voronov, A.~Dzardanov, C.~Williams, and R.~Sobolewski,
  \enquote{{Picosecond superconducting single-photon optical detector},}
  {\protect\JournalTitle{Applied Physics Letters}} \textbf{79}, 705--707
  (2001).

\bibitem{Cabrera1998}
B.~Cabrera, R.~M. Clarke, P.~Colling, A.~J. Miller, S.~Nam, and R.~W. Romani,
  \enquote{{Detection of single infrared, optical, and ultraviolet photons
  using superconducting transition edge sensors},}
  {\protect\JournalTitle{Applied Physics Letters}} \textbf{73}, 735--737
  (1998).

\bibitem{Kardyna2008}
B.~E. Kardyna{\l}, Z.~L. Yuan, and A.~J. Shields, \enquote{{An
  avalanche‐photodiode-based photon-number-resolving detector},}
  {\protect\JournalTitle{Nature Photonics}} \textbf{2}, 425--428 (2008).

\bibitem{Gerrits2010}
T.~Gerrits, S.~Glancy, T.~S. Clement, B.~Calkins, A.~E. Lita, A.~J. Miller,
  A.~L. Migdall, S.~W. Nam, R.~P. Mirin, and E.~Knill, \enquote{{Generation of
  optical coherent-state superpositions by number-resolved photon subtraction
  from the squeezed vacuum},} {\protect\JournalTitle{Physical Review A}}
  \textbf{82}, 031802 (2010).

\bibitem{Namekata2010}
N.~Namekata, Y.~Takahashi, G.~Fujii, D.~Fukuda, S.~Kurimura, and S.~Inoue,
  \enquote{{Non-Gaussian operation based on photon subtraction using a
  photon-number-resolving detector at a telecommunications wavelength},}
  {\protect\JournalTitle{Nature Photonics}} \textbf{4}, 655--660 (2010).

\bibitem{Giustina2015}
M.~Giustina, M.~A. Versteegh, S.~Wengerowsky, J.~Handsteiner, A.~Hochrainer,
  K.~Phelan, F.~Steinlechner, J.~Kofler, J.~{\AA}. Larsson, C.~Abell{\'{a}}n,
  W.~Amaya, V.~Pruneri, M.~W. Mitchell, J.~Beyer, T.~Gerrits, A.~E. Lita, L.~K.
  Shalm, S.~W. Nam, T.~Scheidl, R.~Ursin, B.~Wittmann, and A.~Zeilinger,
  \enquote{{Significant-Loophole-Free Test of Bell's Theorem with Entangled
  Photons},} {\protect\JournalTitle{Physical Review Letters}} \textbf{115},
  250401 (2015).

\bibitem{Niwa2017}
K.~Niwa, T.~Numata, K.~Hattori, and D.~Fukuda, \enquote{{Few-photon color
  imaging using energy-dispersive superconducting transition-edge sensor
  spectrometry},} {\protect\JournalTitle{Scientific Reports}} \textbf{7}, 45660
  (2017).

\bibitem{Zhong2020}
H.-S. Zhong, H.~Wang, Y.-H. Deng, M.-C. Chen, L.-C. Peng, Y.-H. Luo, J.~Qin,
  D.~Wu, X.~Ding, Y.~Hu, P.~Hu, X.-Y. Yang, W.-J. Zhang, H.~Li, Y.~Li,
  X.~Jiang, L.~Gan, G.~Yang, L.~You, Z.~Wang, L.~Li, N.-L. Liu, C.-Y. Lu, and
  J.-W. Pan, \enquote{{Quantum computational advantage using photons},}
  {\protect\JournalTitle{Science}} \textbf{8770}, 1460--1463 (2020).

\bibitem{Lloyd1999}
S.~Lloyd and S.~L. Braunstein, \enquote{{Quantum computation over continuous
  variables},} {\protect\JournalTitle{Physical Review Letters}} \textbf{82},
  1784--1787 (1999).

\bibitem{Andersen2015}
U.~L. Andersen, J.~S. Neergaard-Nielsen, P.~van Loock, and A.~Furusawa,
  \enquote{{Hybrid discrete- and continuous-variable quantum information},}
  {\protect\JournalTitle{Nature Physics}} \textbf{11}, 713--719 (2015).

\bibitem{Takeda2019}
S.~Takeda and A.~Furusawa, \enquote{{Toward large-scale fault-tolerant
  universal photonic quantum computing},} {\protect\JournalTitle{APL
  Photonics}} \textbf{4}, 060902 (2019).

\bibitem{Gottesman2001}
D.~Gottesman, A.~Kitaev, and J.~Preskill, \enquote{{Encoding a qubit in an
  oscillator},} {\protect\JournalTitle{Physical Review A. Atomic, Molecular,
  and Optical Physics}} \textbf{64}, 123101--1231021 (2001).

\bibitem{Alexander2018}
R.~N. Alexander, S.~Yokoyama, A.~Furusawa, and N.~C. Menicucci,
  \enquote{{Universal quantum computation with temporal-mode bilayer square
  lattices},} {\protect\JournalTitle{Physical Review A}} \textbf{97}, 032302
  (2018).

\bibitem{Yukawa2013}
M.~Yukawa, K.~Miyata, T.~Mizuta, H.~Yonezawa, P.~Marek, R.~Filip, and
  A.~Furusawa, \enquote{{Generating superposition of up-to three photons for
  continuous variable quantum information processing},}
  {\protect\JournalTitle{Optics Express}} \textbf{21}, 5529--5535 (2013).

\bibitem{Asavanant2017}
W.~Asavanant, K.~Nakashima, Y.~Shiozawa, J.-I. Yoshikawa, and A.~Furusawa,
  \enquote{{Generation of highly pure Schr{\"{o}}dinger's cat states and
  real-time quadrature measurements via optical filtering},}
  {\protect\JournalTitle{Optics Express}} \textbf{25}, 32227--32242 (2017).

\bibitem{Ra2020a}
Y.-S. Ra, A.~Dufour, M.~Walschaers, C.~Jacquard, T.~Michel, C.~Fabre, and
  N.~Treps, \enquote{{Non-Gaussian quantum states of a multimode light field},}
  {\protect\JournalTitle{Nature Physics}} \textbf{16}, 144--147 (2020).

\bibitem{Jiang2007}
L.~A. Jiang, E.~A. Dauler, and J.~T. Chang, \enquote{{Photon-number-resolving
  detector with 10 bits of resolution},} {\protect\JournalTitle{Physical Review
  A}} \textbf{75}, 2--6 (2007).

\bibitem{Divochiy2008}
A.~Divochiy, F.~Marsili, D.~Bitauld, A.~Gaggero, R.~Leoni, F.~Mattioli,
  A.~Korneev, V.~Seleznev, N.~Kaurova, O.~Minaeva, G.~Gol'tsman, K.~G.
  Lagoudakis, M.~Benkhaoul, F.~L{\'{e}}vy, and A.~Fiore,
  \enquote{{Superconducting nanowire photon-number-resolving detector at
  telecommunication wavelengths},} {\protect\JournalTitle{Nature Photonics}}
  \textbf{2}, 302--306 (2008).

\bibitem{Natarajan2013}
C.~M. Natarajan, L.~Zhang, H.~Coldenstrodt-Ronge, G.~Donati, S.~N. Dorenbos,
  V.~Zwiller, I.~A. Walmsley, and R.~H. Hadfield, \enquote{{Quantum detector
  tomography of a time-multiplexed superconducting nanowire single-photon
  detector at telecom wavelengths},} {\protect\JournalTitle{Optics Express}}
  \textbf{21}, 893--902 (2013).

\bibitem{Gaggero2018}
A.~Gaggero, F.~Martini, F.~Mattioli, F.~Chiarello, R.~Cernansky, A.~Politi, and
  R.~Leoni, \enquote{{Amplitude-multiplexed readout of single photon detectors
  based on superconducting nanowires},} {\protect\JournalTitle{Optica}}
  \textbf{6}, 823--828 (2019).

\bibitem{Yabuno2020}
M.~Yabuno, S.~Miyajima, S.~Miki, and H.~Terai, \enquote{{Scalable
  implementation of a superconducting nanowire single-photon detector array
  with a superconducting digital signal processor},}
  {\protect\JournalTitle{Optics Express}} \textbf{28}, 12047--12057 (2020).

\bibitem{Fukuda2011b}
D.~Fukuda, G.~Fujii, T.~Numata, K.~Amemiya, A.~Yoshizawa, H.~Tsuchida,
  H.~Fujino, H.~Ishii, T.~Itatani, S.~Inoue, and T.~Zama,
  \enquote{{Titanium-based transition-edge photon number resolving detector
  with 98{\%} detection efficiency with index-matched small-gap fiber
  coupling.}} {\protect\JournalTitle{Optics Express}} \textbf{19}, 870--875
  (2011).

\bibitem{Lita2008}
A.~E. Lita, A.~J. Miller, and S.~W. Nam, \enquote{{Counting near-infrared
  single-photons with 95{\%} efficiency},} {\protect\JournalTitle{Optics
  Express}} \textbf{16}, 3032--3040 (2008).

\bibitem{Sperling2012}
J.~Sperling, W.~Vogel, and G.~S. Agarwal, \enquote{{True photocounting
  statistics of multiple on-off detectors},} {\protect\JournalTitle{Physical
  Review A}} \textbf{85}, 023820 (2012).

\bibitem{Jonsson2019}
M.~J{\"{o}}nsson and G.~Bj{\"{o}}rk, \enquote{{Evaluating the performance of
  photon-number-resolving detectors},} {\protect\JournalTitle{Physical Review
  A}} \textbf{99}, 043822 (2019).

\bibitem{Bartlett2002}
S.~D. Bartlett and B.~C. Sanders, \enquote{{Universal continuous-variable
  quantum computation: Requirement of optical nonlinearity for photon
  counting},} {\protect\JournalTitle{Physical Review A - Atomic, Molecular, and
  Optical Physics}} \textbf{65}, 042304 (2002).

\bibitem{Provaznik2020}
J.~Provazn{\'{i}}k, L.~Lachman, R.~Filip, and P.~Marek, \enquote{{Benchmarking
  photon number resolving detectors},} {\protect\JournalTitle{Optics Express}}
  \textbf{28}, 14839--14849 (2020).

\bibitem{Kruse2017}
R.~Kruse, J.~Tiedau, T.~J. Bartley, S.~Barkhofen, and C.~Silberhorn,
  \enquote{{Limits of the time-multiplexed photon-counting method},}
  {\protect\JournalTitle{Physical Review A}} \textbf{95}, 023815 (2017).

\bibitem{Reddy2020}
D.~V. Reddy, R.~R. Nerem, S.~W. Nam, R.~P. Mirin, and V.~B. Verma,
  \enquote{{Superconducting nanowire single-photon detectors with 98{\%} system
  detection efficiency at 1550 nm},} {\protect\JournalTitle{Optica}}
  \textbf{7}, 1649--1653 (2020).

\bibitem{Korzh2020}
B.~Korzh, Q.-Y. Zhao, J.~P. Allmaras, S.~Frasca, T.~M. Autry, E.~A. Bersin,
  A.~D. Beyer, R.~M. Briggs, B.~Bumble, M.~Colangelo, G.~M. Crouch, A.~E. Dane,
  T.~Gerrits, A.~E. Lita, F.~Marsili, G.~Moody, C.~Pe{\~{n}}a, E.~Ramirez,
  J.~D. Rezac, N.~Sinclair, M.~J. Stevens, A.~E. Velasco, V.~B. Verma, E.~E.
  Wollman, S.~Xie, D.~Zhu, P.~D. Hale, M.~Spiropulu, K.~L. Silverman, R.~P.
  Mirin, S.~W. Nam, A.~G. Kozorezov, M.~D. Shaw, and K.~K. Berggren,
  \enquote{{Demonstration of sub-3 ps temporal resolution with a
  superconducting nanowire single-photon detector},}
  {\protect\JournalTitle{Nature Photonics}} pp. 250--255 (2020).

\bibitem{Lamas-Linares2013}
A.~Lamas-Linares, B.~Calkins, N.~A. Tomlin, T.~Gerrits, A.~E. Lita, J.~Beyer,
  R.~P. Mirin, and S.~{Woo Nam}, \enquote{{Nanosecond-scale timing jitter for
  single photon detection in transition edge sensors},}
  {\protect\JournalTitle{Applied Physics Letters}} \textbf{102}, 231117 (2013).

\bibitem{Cahall2017}
C.~Cahall, K.~L. Nicolich, N.~T. Islam, G.~P. Lafyatis, A.~J. Miller, D.~J.
  Gauthier, and J.~Kim, \enquote{{Multi-photon detection using a conventional
  superconducting nanowire single-photon detector},}
  {\protect\JournalTitle{Optica}} \textbf{4}, 1534--1535 (2017).

\bibitem{Zhu2020}
D.~Zhu, M.~Colangelo, C.~Chen, B.~A. Korzh, F.~N. Wong, M.~D. Shaw, and K.~K.
  Berggren, \enquote{{Resolving photon numbers using a superconducting nanowire
  with impedance-matching taper},} {\protect\JournalTitle{Nano Letters}}
  \textbf{20}, 3858--3863 (2020).

\bibitem{Lundeen2009}
J.~S. Lundeen, A.~Feito, H.~Coldenstrodt-Ronge, K.~L. Pregnell, C.~Silberhorn,
  T.~C. Ralph, J.~Eisert, M.~B. Plenio, and I.~A. Walmsley,
  \enquote{{Tomography of quantum detectors},} {\protect\JournalTitle{Nature
  Physics}} \textbf{5}, 27--30 (2009).

\bibitem{Natarajan2012a}
C.~M. Natarajan, M.~G. Tanner, and R.~H. Hadfield, \enquote{{Superconducting
  nanowire single-photon detectors: Physics and applications},}
  {\protect\JournalTitle{Superconductor Science and Technology}} \textbf{25},
  063001 (2012).

\bibitem{Renema2014a}
J.~J. Renema, R.~Gaudio, Q.~Wang, Z.~Zhou, A.~Gaggero, F.~Mattioli, R.~Leoni,
  D.~Sahin, M.~J. {De Dood}, A.~Fiore, and M.~P. {Van Exter},
  \enquote{{Experimental test of theories of the detection mechanism in a
  nanowire superconducting single photon detector},}
  {\protect\JournalTitle{Physical Review Letters}} \textbf{112}, 117604 (2014).

\bibitem{Bulaevskii2012}
L.~N. Bulaevskii, M.~J. Graf, and V.~G. Kogan, \enquote{{Vortex-assisted photon
  counts and their magnetic field dependence in single-photon superconducting
  detectors},} {\protect\JournalTitle{Physical Review B}} \textbf{85}, 014505
  (2012).

\bibitem{Heeres2012}
R.~W. Heeres and V.~Zwiller, \enquote{{Superconducting detector dynamics
  studied by quantum pump-probe spectroscopy},} {\protect\JournalTitle{Applied
  Physics Letters}} \textbf{101}, 112603 (2012).

\bibitem{Marsili2016}
F.~Marsili, M.~J. Stevens, A.~Kozorezov, V.~B. Verma, C.~Lambert, J.~A. Stern,
  R.~D. Horansky, S.~Dyer, S.~Duff, D.~P. Pappas, A.~E. Lita, M.~D. Shaw, R.~P.
  Mirin, and S.~W. Nam, \enquote{{Hotspot relaxation dynamics in a
  current-carrying superconductor},} {\protect\JournalTitle{Physical Review B}}
  \textbf{93}, 094518 (2016).

\bibitem{Nicolich2019}
K.~L. Nicolich, C.~Cahall, N.~T. Islam, G.~P. Lafyatis, J.~Kim, A.~J. Miller,
  and D.~J. Gauthier, \enquote{{Universal Model for the Turn-On Dynamics of
  Superconducting Nanowire Single-Photon Detectors},}
  {\protect\JournalTitle{Physical Review Applied}} \textbf{12}, 034020 (2019).

\bibitem{Yamashita2016}
T.~Yamashita, K.~Waki, S.~Miki, R.~A. Kirkwood, R.~H. Hadfield, and H.~Terai,
  \enquote{{Superconducting nanowire single-photon detectors with non-periodic
  dielectric multilayers},} {\protect\JournalTitle{Scientific Reports}}
  \textbf{6}, 35240 (2016).

\bibitem{Miki2013}
S.~Miki, T.~Yamashita, H.~Terai, and Z.~Wang, \enquote{{High performance
  fiber-coupled NbTiN superconducting nanowire single photon detectors with
  Gifford-McMahon cryocooler},} {\protect\JournalTitle{Optics Express}}
  \textbf{21}, 10208--10214 (2013).

\bibitem{Miki2010}
S.~Miki, T.~Yamashita, M.~Fujiwara, M.~Sasaki, and Z.~Wang,
  \enquote{{Multichannel SNSPD system with high detection efficiency at
  telecommunication wavelength},} {\protect\JournalTitle{Optics Letters}}
  \textbf{35}, 2133--2135 (2010).

\bibitem{Kerman2013}
A.~J. Kerman, D.~Rosenberg, R.~J. Molnar, and E.~A. Dauler, \enquote{{Readout
  of superconducting nanowire single-photon detectors at high count rates},}
  {\protect\JournalTitle{Journal of Applied Physics}} \textbf{113}, 144511
  (2013).

\bibitem{Cahall2018}
C.~Cahall, D.~J. Gauthier, and J.~Kim, \enquote{{Scalable cryogenic readout
  circuit for a superconducting nanowire single-photon detector system},}
  {\protect\JournalTitle{Review of Scientific Instruments}} \textbf{89}, 063117
  (2018).

\bibitem{DAriano2004}
G.~M. D'Ariano, L.~Maccone, and P.~L. Presti, \enquote{{Quantum calibration of
  measurement instrumentation},} {\protect\JournalTitle{Physical Review
  Letters}} \textbf{93}, 250407 (2004).

\bibitem{Luis1999}
A.~Luis and L.~L. S{\'{a}}nchez-Soto, \enquote{{Complete characterization of
  arbitrary quantum measurement processes},} {\protect\JournalTitle{Physical
  Review Letters}} \textbf{83}, 3573--3576 (1999).

\bibitem{Yokoyama2019}
S.~Yokoyama, N.~{Dalla Pozza}, T.~Serikawa, K.~B. Kuntz, T.~A. Wheatley,
  D.~Dong, E.~H. Huntington, and H.~Yonezawa, \enquote{{Characterization of
  entangling properties of quantum measurement via two-mode quantum detector
  tomography using coherent state probes},} {\protect\JournalTitle{Optics
  Express}} \textbf{27}, 34416--34432 (2019).

\bibitem{Achilles2004}
D.~Achilles, C.~Silberhorn, C.~Sliwa, K.~Banaszek, I.~A. Walmsley, M.~J. Fitch,
  B.~C. Jacobs, T.~B. Pittman, and J.~D. Franson,
  \enquote{{Photon-number-resolving detection using time-multiplexing},}
  {\protect\JournalTitle{Journal of Modern Optics}} \textbf{51-9}, 1499--1515
  (2004).

\end{thebibliography}






\end{document}